# Lab-Tutorials für den Quantenphysik Unterricht

Michael C. Wittmann, Department of Physics and Astronomy, University of Maine,
Orono ME 04469-5709, wittmann@umit.maine.edu, http://perlnet.umaine.edu/

## Zusammenfassung

Im Kurs „Intuitive Quantum Physics“ werden graphische Interpretationen mathematischer Gleichungen und qualitatives Denken durch ein vereinfachtes Modell der Quantenphysik gelehrt. Unser Kurs besteht aus drei wichtigen Abschnitten: Wellenphysik, Aufbau eines Werkzeugkastens („Toolbox“) und Quantenphysik, sowie drei Schlüsselthemen: Welle-Teilchen-Dualität, die Schrödinger-Gleichung und Tunneln von Quantenteilchen. Wir unterrichten vorwiegend mit Lab-Tutorials, in denen StudentInnen in kleinen Gruppen (3 bis 4 Personen) anwendungsspezifische Arbeitsblätter durcharbeiten. In den Diskussionen werden auch Auseinandersetzungen über das „Bild der Physik“, bei uns „Nature of Science“ genannt, geführt. Überprüfungen haben ergeben, dass StudentInnen nicht nur die schwierigsten Konzepte des Kurses lernen können sondern auch lernen, dass die Quantenphysik begrifflich verständlich ist. Diese Einsicht ist besonders für zukünftige Lehrkräfte von Bedeutung.

## Einleitung

Nach mehreren Jahren Erfahrung mit diesem Kurs „Intuitive Quantum Physics“ (IQP) sind wir überzeugt, dass StudentInnen die wichtigen Konzepte der Quantenphysik lernen können ohne anspruchsvolle Mathematik zu benutzen, also vorwiegend durch qualitatives Denken und einfache Analysen von graphischen Darstellungen. Obwohl der Kurs in einer Universität in den Vereinigten Staaten unterrichtet wird, kann man ohne weiteres die StudentInnen in diesem Kurs an der University of Maine mit SchülerInnen der Oberstufe in Deutschland vergleichen. Genauer gesagt, StudentInnen an der Universität Maine sind wahrscheinlich schwächer in der Mathematik als Oberstufen-SchülerInnen in Physikklassen in den Gymnasien Deutschlands und Österreichs. Daher ist unser IQP Kurs auf einer sehr konzeptuellen und praktischen Basis aufgebaut. In diesem Aufsatz beschreiben wir unsere Ziele im Unterricht, die drei wichtigen Abschnitte des Kurses und auch die Belege dafür, dass der IQP Kurs erfolgreich ist.

Unser Ziel ist nicht nur, dass StudentInnen nach 5 Jahren die Konzepte der Physik verstehen und sich daran erinnern, sondern viel mehr, dass sie lernen, dass auch in der anscheinend widerspruchsvollen Quantenwelt eine verständliche und zusammenhängende Beschreibung der Natur möglich ist.

Der Kurs besteht aus drei Teilen:

- Einführung in die Wellenphysik,
- Aufbau eines Werkzeugkastens („toolbox“) mit wichtigen Konzepten, die der Interpretationen der Schrödinger-Gleichung dienen (darunter Energie, Wahrscheinlichkeit und Krümmung), und
- Anwendung des Werkzeugkastens auf einfache Modelle eines Atoms, eines Moleküls und der radioaktiven Strahlung.

Wir messen den Erfolg im IQP-Kurs mit mehreren Methoden, beschreiben hier aber nur zwei: Fragebögen zu Einstellungen zur Physik und den Naturwissenschaften [1,2] und Tests über das Tunneln von Quantenteilchen [3]. Wir haben noch weitere Daten, aber die genannten zeigen am deutlichsten, dass StudentInnen mit sehr schwierigen Konzepten umgehen können und dass sie sich bewusst sind, ein konzeptuelles Bild der Physik aufzubauen. Beides ist uns wichtig, aber ein besseres Verständnis der Naturwissenschaft als selbst konstruiertes Wissen ist uns im Endeffekt noch wichtiger: Es ist die Basis für weiteres Lernen.

Wir verzichten hier darauf, die University of Maine, die StudentInnen und die Unterrichtsmethoden im Detail zu beschreiben. Viel wichtiger ist es für uns, einen Überblick über die wichtigen Konzepte zu geben und einige Beispiele der praktischen Tätigkeiten im Klassenzimmer zu beschreiben. Pro Woche haben wir 6 Stunden Unterricht. Drei Stunden sind Vorlesung mit ca. 50 StudentInnen, drei Stunden sind „Lab-Tutorials“ im Sinne der „Tutorials in Introductory Physics“ die an der University of Washington entwickelt wurden [4]. Die IQP-Lab-Tutorials sind zum Großteil in der Uni Maine von Mitgliedern des Physics Education Research Laboratory entwickelt. Details über Lab-Tutorials werden später beschrieben.

## Welche Physik ist wichtig?

Auch für erfahrene Physiker ist die Quantenphysik nicht einfach zu verstehen. Es ist relativ leicht, die Mathematik zu erlernen, aber auch die einfachsten Konzepte bleiben schwer verständlich. Man betrachte zum Beispiel was passiert, wenn einzelne Elektronen durch einen Doppelspalt geschossen werden. Wartet man lange genug, sieht man das folgende Bild:

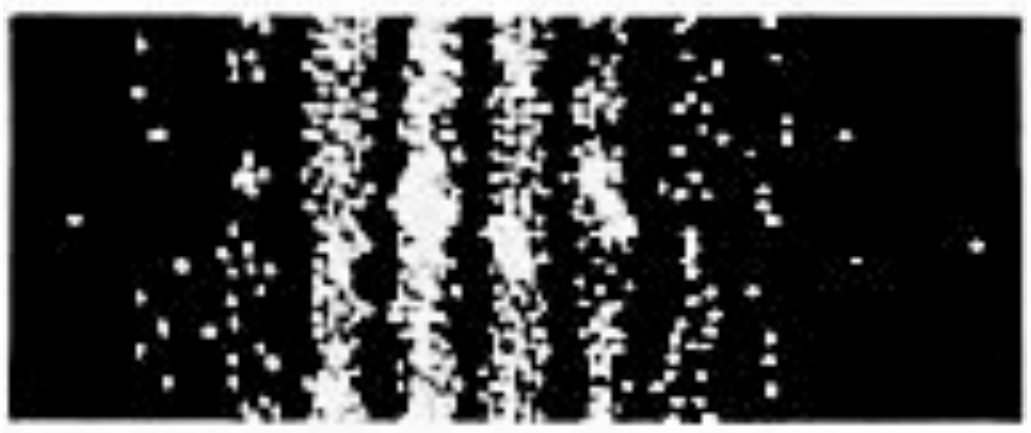

**Abbildung 1:** Elektroneninterferenz für das Doppelspalt-Experiment [5]

Die weißen Punkte stellen die Elektronen dar, die auf der Platte landen. Wie soll man das verstehen? Zumindest zwei Analogien sind möglich:

1. Die weißen Punkte am Bild sind wie Flecken auf einer Wand, gegen die nasse Tennisbälle geworfen werden.
2. Dunkle Stellen sind wie die Interferenzminima aus der physikalischen Optik, wo zum Beispiel Interferenz von Laserstrahlen zu einem ähnlichen Bild führt.

Das Problem mit diesen Analogien liegt darin, dass beide nicht zugleich stimmen können. Man kann zwei logische Widersprüche finden, sowohl in der Wellen- als auch in der Teilchen-Interpretation.

a. Ein Teilchen das wir als Punkt im Bild sehen, muss durch beide Spalte gehen, um Interferenz zu verursachen. Teilchen können aber nur durch einen Spalt hindurch fliegen.
b. Durch der Beobachtung der Interferenz folgern wir, dass es sich um Wellen handelt die durch beide Spalten gehen, jedoch als Punkt am Bild zu sehen sind. Wellen sind aber nicht lokalisierbar.

Um diese Widersprüche zu verstehen, müssen StudentInnen (und LehrerInnen und ProfessorInnen natürlich auch) eine Fülle an Physik wissen, unter anderem: Wellenphysik, Superposition, Interferenz, Energie, Wahrscheinlichkeit und den Unterschied zwischen begrenzten, lokalisierten Teilchen und unbegrenzten Wellen. Wir haben also unseren IQP-Kurs aufgebaut um das Ziel erreichen zu können, diese Konzepte den StudentInnen beizubringen und durch Übung in ein selbst konstruiertes gedankliches Modell der Quantenphysik zu integrieren.

## Beschreibung des Kurses

Die drei Abschnitte des Kurses sind in Tabelle 1 beschrieben. In diesem Aufsatz wollen wir vorwiegend über die Ideen des zweiten und dritten Abschnittes berichten, denn in diesen Abschnitten werden die Ideen der Quantenphysik zuerst formuliert und angewandt. Wir geben Beispiele von Lab-Tutorials in denen StudentInnen zum ersten Mal mit den wichtigsten Ideen des Kurses bekannt gemacht werden. Falls notwendig ist, beschreiben wir Vorlesungsmethoden die mit den Lab-Tutorials verbunden sind.

| | Inhalt | Methoden der Naturwissenschaft |
|---|---|---|
| 1. Abschnitt: Optik und Wellenphysik | • Licht geht von jedem Punkt gerade aus in alle Richtungen<br>• Superposition von Wasser- und Lichtwellen<br>• Interferenz bei Wasser und Licht<br>• Wellen-Teilchen-Dualität – das große Dilemma! | • Woher wissen Sie das? „Ich habe es gesehen“<br>• Wie erklären Sie das? „Es ist wie dieses andere Objekt“<br>• Warum glauben Sie es? „Weil ich es gesehen habe“ |
| 2. Abschnitt: Werkzeugkasten für Quantenphysik | • Energiedarstellungen<br>• Klassische Wahrscheinlichkeit<br>• Krümmungen von Funktionen in graphischen Darstellungen<br>• Eine graphische Interpretation der Schrödinger-Gleichung | • Woher wissen Sie das? „Sie haben es mir gesagt“<br>• Wie erklären Sie das? „Es ist wie etwas anderes, das ich kenne.“<br>• Wieso glauben Sie es? „Ich habe darüber nachgedacht“ |
| 3. Abschnitt: Themen der Quantenphysik | • Quantisierung in begrenzten Potentialen; Gebundene Zustände<br>• Spektroskopie<br>• Einfache Modelle des $H_2^+$ -Moleküls<br>• Tunneln von Quantenteilchen | • Woher wissen Sie das? „Ich habe es ausgearbeitet“<br>• Wie erklären Sie das? „Es stimmt mit anderen Ideen überein”<br>• Wieso glauben Sie es? „Ich weiß nicht, ob ich es glaube, ich kann trotzdem darüber nachdenken.“ |

**Tabelle 1:** Kursaufbau für „Intuitive Quantum Physics“

## 1. Abschnitt: Optik und Wellenphysik

Da der erste Abschnitt von Optik und Wellenphysik handelt, beschreiben wir den Inhalt nur kurz und gehen auf den Kursaufbau genauer ein. Die Unterrichtsmaterialien wurden großteils von zwei im Handel erhältlichen Produkten abgeleitet: *Tutorials in Introductory Physics* [4] und *Activity-Based Physics Volume 1: Introductory Physics* [6] und *Volume 2: Modern Physics* [7]. Tutorials wurden speziell für die Arbeit in den Übungsstunden zu einer Universitätsvorlesung entworfen [8-10]. Man arbeitet in kleinen Gruppen (3 bis 4 Personen) an anwendungsorientierten Arbeitsblättern. Ergebnisse von Untersuchungen zum begrifflichen Lernprozess der StudentInnen belegen den Nutzen dieser Arbeitsform. Die Anwendung von Mathematik wird nicht betont. Stattdessen beschreibt man die Physik qualitativ, wie in Abbildung 2 abgebildet wird (Text aus [6], aus dem Englischen übersetzt). In diesem Beispiel werden die Geschwindigkeiten der einzelnen Federteilchen ohne Mathematik miteinander verglichen, d.h. man muss nicht $v_y$ von (dy/dx) (dx/dt) ableiten, sondern kann einfach das Bild selber benutzen um $v_y(B)$ mit $v_y(D)$ zu vergleichen.

Im IQP-Kurs haben wir die bereits existierenden Materialien adaptiert, indem wir den Großteil der formalen mathematischen Betrachtungen in qualitative Überlegungen abänderten. Die Kernideen, die für die Quantenphysik (besser gesagt für die Wellen-Teilchen-Dualität) wichtig sind, dienen alle dazu, das Konzept der Interferenz zu entwickeln. Wir vermeiden mathematische Betrachtungen, zum Beispiel Phasendifferenzen zwischen zwei Wellen, Sinuswellen und ähnliches, wir betonen die qualitative Beschreibung der Beobachtungen. Wird die Welle schneller wenn man die Hand, die eine lange Schnur hält, schneller auf und nieder bewegt [11]? Man kann aus Beobachtungen schließen, dass die Geschwindigkeit einer Welle von der Spannung der Schnur abhängt, aber nicht davon, wie die Welle erzeugt wird.

Die wichtigsten Konzepte werden also nicht mathematisch sondern aus einfachen Beobachtungen und Überlegungen hergeleitet. Wenn notwendig, zum Beispiel bei der Superposition von zwei schnellen Wellen auf einer fest gespannten Schnur, benutzen wir Videos in denen man die Wellen in Zeitlupe anschauen kann [6,11-13]. Wir vermeiden Simulationen, damit die StudentInnen auch beim Modellaufbau immer einen engen Zusammenhang mit der realen Welt haben. Stattdessen benutzen wir Videos, Wasserwannen und lange Sprungfedern, um Wellen zu beobachten. Wir benutzen auch einfache Glühbirnen, Laser und Blenden mit großen Löchern oder kleinen Spalten in der Optik.

Der erste Abschnitt endet mit einer Diskussion der Wellen-Teilchen-Dualität. StudentInnen können sowohl die Interferenz über die Intensitätsbilder als auch die Teilchen-Punkte der Elektronen (und auch Photonen) über die Einzelereignisse auf dem Schirm einsehen. Die Widersprüche werden explizit diskutiert. (Wir geben ein Beispiel des Lab-Tutorials im Internet bei arxiv.org [14] frei.). Das Dilemma wird in der Vorlesung ausführlich diskutiert, sowohl mit allen 50 StudentInnen als auch in kleineren Gruppen von 3 bis 10, die jeweils eine Position vertreten.

## 2. Abschnitt: Ein Wergzeugkasten für die Schrödinger-Gleichung

Um die Wellen-Teilchen Dualität besser zu verstehen, müssen wir einen neuen Werkzeugkasten erstellen. Wir führen die Konzepte von Energie, Wahrscheinlichkeit und Krümmung einer graphischen Kurve ein. Interferenz war das leitende Beispiel des ersten Abschnittes, hier ist es die Notwendigkeit der Einführung der Schrödinger-Gleichung.

Rechts ist ein Wellenpuls auf einer Feder abgebildet. Der Wellenpuls bewegt sich nach rechts mit einer Geschwindigkeit von 100 cm/s. Punkte A-F zeigen wo kleine Schnürchen auf die Feder fest gebunden sind. Jeder Kasten in der horizontalen Richtung entspricht 1 cm.

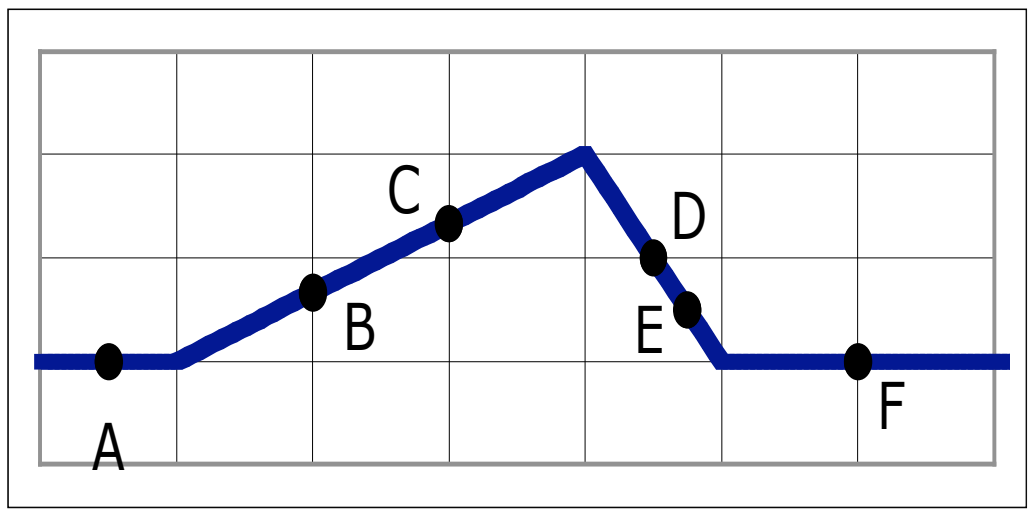

1. Zeichne ins Bild die Feder 0,01 s nach dem abgebildeten Zeitpunkt.
2. In welche Richtung bewegen sich die Schnürchen (z.B. bei Punkt D)? Zeichne auf dem Bild die Richtung mit einem Pfeil ein.
3. Bewegen sich alle Schnürchen auf der Feder in die gleiche Richtung?
4. Vergleiche die Richtung der Bewegung der Schnürchen und des Wellenpulses.
5. Vergleiche die Beschreibung der Schnürchen und Wellenpulses mit den Beobachtungen, mit der wir den Unterricht begannen.

**Abbildung 2:** Qualitative Fragen über die Bewegung eines Wellenpulses auf einer langen Feder

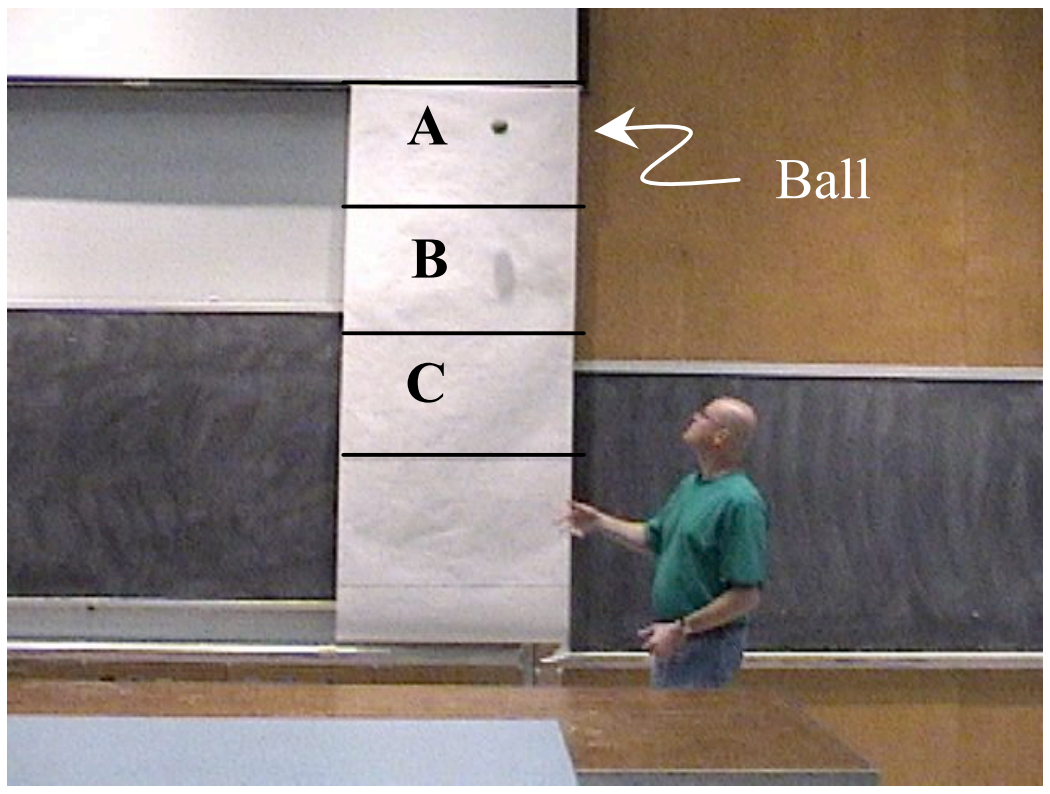

**Abbildung 3:** Flug eines Balles durch drei Zonen

Wahrscheinlichkeit führen wir durch einfache Experimente ein. Wir fangen mit Münzen und Würfeln an. Danach interpretieren StudentInnen den Flug eines Balles, der vertikal in die Luft geworfen wird. Man stelle sich vor, die Flughöhe wird in drei gleich große Zonen geteilt (in Abbildung 3 mit A, B und C bezeichnet). Wenn man während des Fluges hinschaut, wo findet man den Ball am ehesten? Viele werden sagen in C, weil er dort landet. Andere werden sagen in B, weil er dort zweimal vorbei kommt. Richtig ist A, weil der Ball sich dort am langsamsten bewegt und daher am längsten in dieser Zone bleibt. StudentInnen arbeiten wieder mit einem Video.

StudentInnen zählen, wie viele Videoframes den Ball in jeder Zone zeigen und können dadurch die Wahrscheinlichkeit bestimmen. Ergebnisse werden sowohl numerisch als auch graphisch mit einem Balkendiagramm gezeigt. Mit ähnlichen Methoden zeigt ein zweites Video graphisch die Wahrscheinlichkeit des harmonischen Oszillators. (Videos sind auf unserer Webseite erhältlich: http://perlnet.umaine.edu/materials/.)

Um den Begriff Energie einzuführen kehren wir zu einfachen Beispielen aus der Kinematik zurück. Den Begriff der kinetischen Energie („Bewegungsenergie“) führen wir ein, indem wir die intuitiven Vorstellungen der StudentInnen am Beispiel eines fallenden Balles erfragen. Ein Ball nimmt „Bewegungsenergie“ auf, wenn er fällt. Das heißt, dass er das Potential hat diese Energie aufzunehmen bevor er fällt. Um potentielle Energie und kinetische

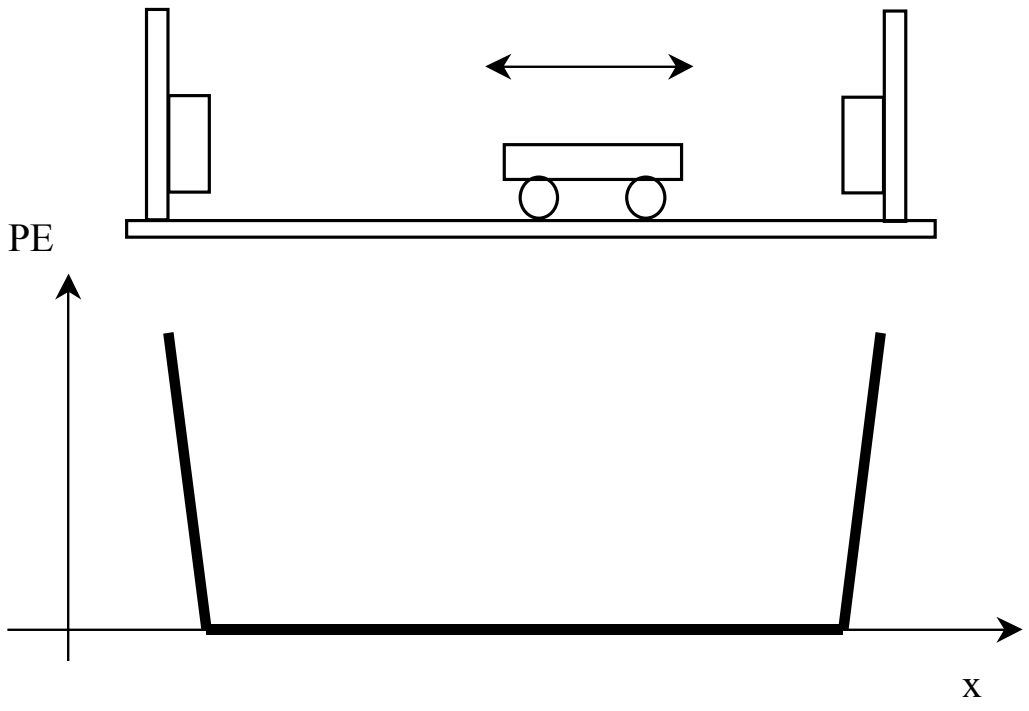

**Abbildung 4**: Fast-quadratisches Potential durch Gedankenexperiment entwickelt

Energie genauer zu beschreiben arbeiten die StudentInnen mit computer-basierten Messgeräten (von Pasco) [15]. Sie untersuchen zuerst ein Auto, das eine Rampe hinauf und herunter rollt und dann eine Masse, die an einer Sprungfeder hängt. Hier wird von ihnen wieder erwartet, dass sie ihre Intuitionen über Energieerhaltung anwenden. Eine wichtige Idee wird hier eingeführt, nämlich dass man einerseits die Energie eines Systems, andererseits die Energie eines Objektes in dem System beschreiben kann. Beide sind wichtig in unserer Formulierung der Schrödinger-Gleichung.

Die praktischen Beobachtungen werden danach in abstrakten Situationen durch Gedankenexperimente weiter entwickelt, zum Beispiel bei der Beschreibung der Energie eines Wagens zwischen zwei Wänden (siehe Abbildung 4). Hier können wir schwer mit dem Computer arbeiten, aber die StudentInnen können leicht das ideale System beschreiben, in dem der Wagen mit konstanter Geschwindigkeit bis gegen die Wand rollt und (ohne Energieverlust) umdreht. Das Ganze können wir graphisch zeigen (siehe Abbildung 4). Dieses Bild eines quadratischen Potentials benutzen wir später wieder, wenn wir über einfache Potentiale in der Quantenphysik reden.

Um die Schrödinger-Gleichung graphisch zu interpretieren, müssen wir noch das Konzept der Krümmung einführen. StudentInnen mit Erfahrung in Differentialrechnung ist das Konzept bekannt. Aber auch für diese StudentInnen benutzen wir eine eher intuitive Formulierung. Die Frage ist, wie sehr man das Lenkrad an einem Auto drehen muss, um eine Kurve zu durchfahren. Je weiter man dreht, umso größer ist die Krümmung der Kurve. Wir fangen mit Kreisen an und betrachten danach Sinuswellen deren Krümmung sich mit dem Wert der Funktion ändert.

Jetzt sind die StudentInnen soweit, dass sie die Schrödinger-Gleichung benutzen können. Wir führen diese zuerst in der Vorlesung ein und üben dann im dritten Kursabschnitt mit Beispielen aus den Lab-Tutorials.

Die zeitunabhängige Schrödinger-Gleichung in einer Dimension,

$$-\frac{\hbar^2}{2m}\frac{d^2\Psi}{dx^2}+U\Psi=E\Psi$$

formulieren wir als

$$\text{Krümmung von } \Psi = -k\left(E_{Teilchen}-E_{System}\right)\Psi$$

wo $E_{Teilchen}$ die Gesamtenergie des Teilchens ist (die im ganzen System gleich ist) und $E_{System}$ die potentielle Energie (als Funktion von $x$) ist. Alle Konstanten haben wir zu $k$ vereinfacht. Eine positive Krümmung für negative Werte der Wellenfunktion $\Psi$ (oder negative Krümmung für positive Werte) ergibt eine Sinuswelle. Dies passiert, wenn die Teilchenenergie in einem Bereich größer ist als die potentielle Energie. Eine positive Krümmung für positive Werte der Wellenfunktion (oder negative Krümmung für negative Werte der Wellenfunktion) ergibt eine Exponentialkurve. Dies passiert, wenn die Teilchenenergie in einem Gebiet kleiner ist als die potentielle Energie des Systems, mit anderen Worten wenn ein Teilchen in einem in der klassischen Physik „unerlaubtem“ Gebiet ist. Die graphische Interpretation der Schrödinger-Gleichung kann man qualitativ beobachten und anwenden.

Um die Schrödinger-Gleichung zu interpretieren, müssen die StudentInnen in ihren Werkzeugkasten greifen. Der Vergleich zwischen System- und Teilchen-Energie bestimmt das Vorzeichen der Krümmung. Wir benutzen am Anfang ein Computerprogramm, um die Wellenfunktion zu zeichnen. Bestimmte Regeln werden benutzt

a) der Wert der Wellenfunktion muss Null sein wenn $x$ unendlich klein oder groß ist,

b) die Wellenfunktion muss bei einem Übergang von einem Gebiet in ein anderes im Wert und der Steigung gleich bleiben, auch wenn sich die Krümmung ändert.

Wir „nähen“ also stetig differenzierbare Wellenfunktionen zusammen die bei Null anfangen und aufhören, sich aber wie Sinuswellen oder Exponentialfunktionen zeichnen lassen. Damit bildet man zum Beispiel die Wellenfunktion für ein Teilchen in einem quadratischen Potential. Man kann überraschend komplizierte Wellenfunktionen zeichnen, die man analytisch nur schwer beschreiben kann [16]. Die Wahrscheinlichkeit findet man, indem man den Wert der Wellenfunktion quadriert. Die Summe aller Wahrscheinlichkeiten muss natürlich 1 (bzw. 100%) sein.

## 3. Abschnitt: Modelle der Quantenwelt

Als erstes Beispiel der Quantenwelt beschreiben die StudentInnen ein Modell eines Atoms. Das „Atom“ ist eigentlich nur ein quadratisches Potential, aber gerade weil es ein schlechtes Modell ist kann man die Frage der Modellbildung in der Physik gut diskutieren. Ein quadratisches Potential erlaubt nur spezifische Energien. Mit diesem Modell können wir in der Spektroskopie das Absorptionsspektrum von Sternen analysieren. Ein quadratisches Potential erlaubt auch die Möglichkeit der Ionisation. Dieser Begriff ist vielen aus der Chemie bekannt. Dass das quadratische Potential natürlich nicht ein richtiges drei-dimensionales Bild eines Atoms zeigt, kann man auch explizit diskutieren. StudentInnen haben zunächst einfach nicht das mathematische Werkzeug, um in drei Dimensionen zu rechnen und wir sind überzeugt, dass einfache Resultate der Quantenphysik anders besser eingeführt werden können.

Wir können einfache Modelle eines Moleküls bilden, zum Beispiel $H_2^+$, indem wir zwei „Atome“ nahe aneinander bringen. Mit einem Computerprogramm zeigen wir, dass die Grundenergie des Systems kleiner wird je näher die Atome zueinander sind. Wieder besteht die Möglichkeit, dass StudentInnen die Rolle eines unvollständigen Modells diskutieren. Obwohl Atome in der Natur nie übereinander liegen gibt es in unserem Modell keinen Grund, eine Distanz zwischen den Atomen zu behalten.

Unser IQP-Kurs endet mit einer Diskussion zum Tunneln von Quantenteilchen. Wir benutzen die gleichen Methoden, brechen aber unsere erste Regel, dass die Wellenfunktion Null sein muss bei unendlich großen oder kleinen Werten von $x$. Stattdessen sagen wir, dass ganz viele Teilchen mit bestimmter Energie auf ein Potential zukommen, also mit einer Wellenfunktion, die durch eine einzige Sinusfunktion zu beschreiben ist. Was passiert bei der quadratischen Potentialbarriere? Ist es möglich ein Teilchen auf der anderen Seite der Potentialbarriere zu finden? Aus unseren graphischen Regeln folgt eindeutig, dass eine Sinuswelle auf der anderen Seite der Potentialbarriere existiert, dass also auch Teilchen dort zu finden sind. Dieses Modell benutzen wir, um die Abstrahlung zu verstehen, was für unsere StudentInnen interessant ist, weil in unserem Bundesstaat, Maine, Radon in Hauskellern und im Grundwasser ein Problem ist [17].

# Zum Unterrichtserfolg

Wir messen den Erfolg des IQP-Kurses sowohl an den Meinungen der StudentInnen über die Physik als auch über das Verständnis der schwierigen Konzepte.

## *Einstellung zur Physik und den Naturwissenschaften*

Abschnitte zwei und drei sind eindeutig sehr schwer für unsere StudentInnen, aber durch Lab-Tutorials ist es möglich, dass sie die Ideen für sich selbst entwickeln. In den Lab-Tutorials lernen sie also für sich selber, nicht durch Auswendiglernen, sondern durch die Entwicklung von kleinen Ideen zu umfassenderen Konzepten.

Die Vorlesung dient dazu, wichtige Fragen zu betonen und zu diskutieren. Es gibt auch eine Meta-Diskussion, in der die Sprache der Lab-Tutorials benutzt wird, um die Konzeptentwicklung der StudentInnen zu beschreiben. Wir „nähen" also Konzepte zusammen, und versuchen stetig und differenzierbar von einem Bereich in andere zu gehen. (Wir glauben aber nicht, dass der Verstand bei Null anfängt und wollen vermeiden dass der Verstand wieder bei Null aufhört. Dieser Witz, explizit in der Vorlesung besprochen, dient wieder dazu, den StudentInnen zu zeigen, dass jede Analogie eine Grenze hat.)

Wir messen die Einstellungen der StudentInnen gegenüber der Physik durch den Maryland Physics Expectations Survey II (MPEX2) [1,2]. Dieser Fragebogen erfasst, was StudentInnen vom Physikunterricht und von sich selbst im Unterricht erwarten. Zum Beispiel: „Ist die Physik auf einer konzeptuellen oder faktischen Grundlage aufgebaut?" Fragen in der Dimension „Konzept" („Concepts Cluster") messen also nicht, ob die StudentInnen die Konzepte kennen, sondern ob sie glauben es sei wichtig, die Konzepte zu kennen. In der Dimension „Zusammenhang" („Coherence Cluster") geht es darum, ob Ideen von einander abhängen oder ob Ideen selbstständig und von einander unabhängig sind. In der Dimension „Unabhängigkeit" („Independence Cluster") wird erfragt, ob man die Ideen der Physik selbst lernen kann oder ob man immer von Experten abhängt. Die Graphik in Abbildung 5 zeigt unsere Resultate von Vortests (vor dem gesamten Unterricht) und Nachtests (nach dem gesamten Unterricht).

Um diese Graphik zu interpretieren, muss man zuerst die Achsen betrachten. Wir messen den Prozentsatz der Fragen, die günstig oder ungünstig beantwortet werden. Zum Beispiel ist es nicht günstig der Feststellung „Um erfolgreich Physik zu lernen muss ich nur in der Klasse zuhören, mein Buch lesen und Aufgaben lösen" zuzustimmen. Zur Physik zählt mehr als das. Der MPEX2 besteht aus 33 ähnlichen Fragen. Es ist also am besten sich in der oberen linken Ecke zu befinden, wo man nur günstige Antworten gibt. Dort findet man Physik-Experten. Weil die Summe von gültigen, ungültigen, und neutralen Antworten 100% ergeben muss, zeigt unsere Graphik nur einen begrenzten Bereich, in dem die Distanz von einem Datenpunkt zu der Diagonale den Prozentsatz neutraler Antworten angibt.

Allgemein sind die Daten mittelmäßig, was typisch ist für diese Population von StudentInnen. Wir sehen, dass sich die Meinungen und Einstellungen der StudentInnen nur wenig in den Dimensionen „Zusammenhang“ und „Selbstständigkeit“ ändern. Was sehr wichtig ist, ist die starke Verbesserung in der Dimension „Konzept“. StudentInnen entwickeln tatsächlich ein besseres Bild der Physik, nicht durch Auswendiglernen von Fakten, sondern durch die selbstständige Konstruktion eines konzeptuellen Modells.

Diese Daten unterscheiden sich von jenen aus fast allen anderen Klassen, die mit dem MPEX2 untersucht wurden [1,2]. Besonders wenn Quantenphysik gelehrt wird, findet man üblicherweise eine starke Verschlechterung der Antworten. Die Daten bewegen sich in die rechte untere Ecke in der Graphik. In anderen Worten bringt unser Kurs den StudentInnen etwas über die konzeptuelle Basis der Physik bei, schadet ihnen aber nicht in ihren Meinungen über Selbständigkeit und Zusammenhang der Ideen der Physik. Diese Resultate sind sehr positiv im Vergleich mit typischen Resultaten von anderen Klassen.

## Konzeptuelles Wissen über Tunneln von Quantenteilchen

Es ist uns natürlich wichtig, dass StudentInnen glauben, dass sie physikalische Konzepte kennen sollen. Es ist auch wichtig, dass sie die Konzepte erlernen. Wir messen unseren Erfolg an der schwierigsten Idee des IQP-Kurses, nämlich das Tunneln von Quantenteilchen. Um unsere Fragen richtig zu beantworten, müssen StudentInnen die Konzepte der Energie und Wahrscheinlichkeit kennen und müssen mit der graphischen Interpretation der Schrödinger-Gleichung umgehen können. Von unserer Arbeit in der Physikdidaktik haben wir gelernt, dass unsere StudentInnen typischerweise das Konzept Wahrscheinlichkeit sehr gut verstehen. Dies steht im Gegenteil zu Resultaten von Domert u.a. [18], aber unsere Ergebnisse betreffen nur sehr einfache Beispiele. StudentInnen sagen aber sehr häufig, dass Teilchen beim Tunneln Energie verlieren. Nur 40% der StudentInnen mit dem

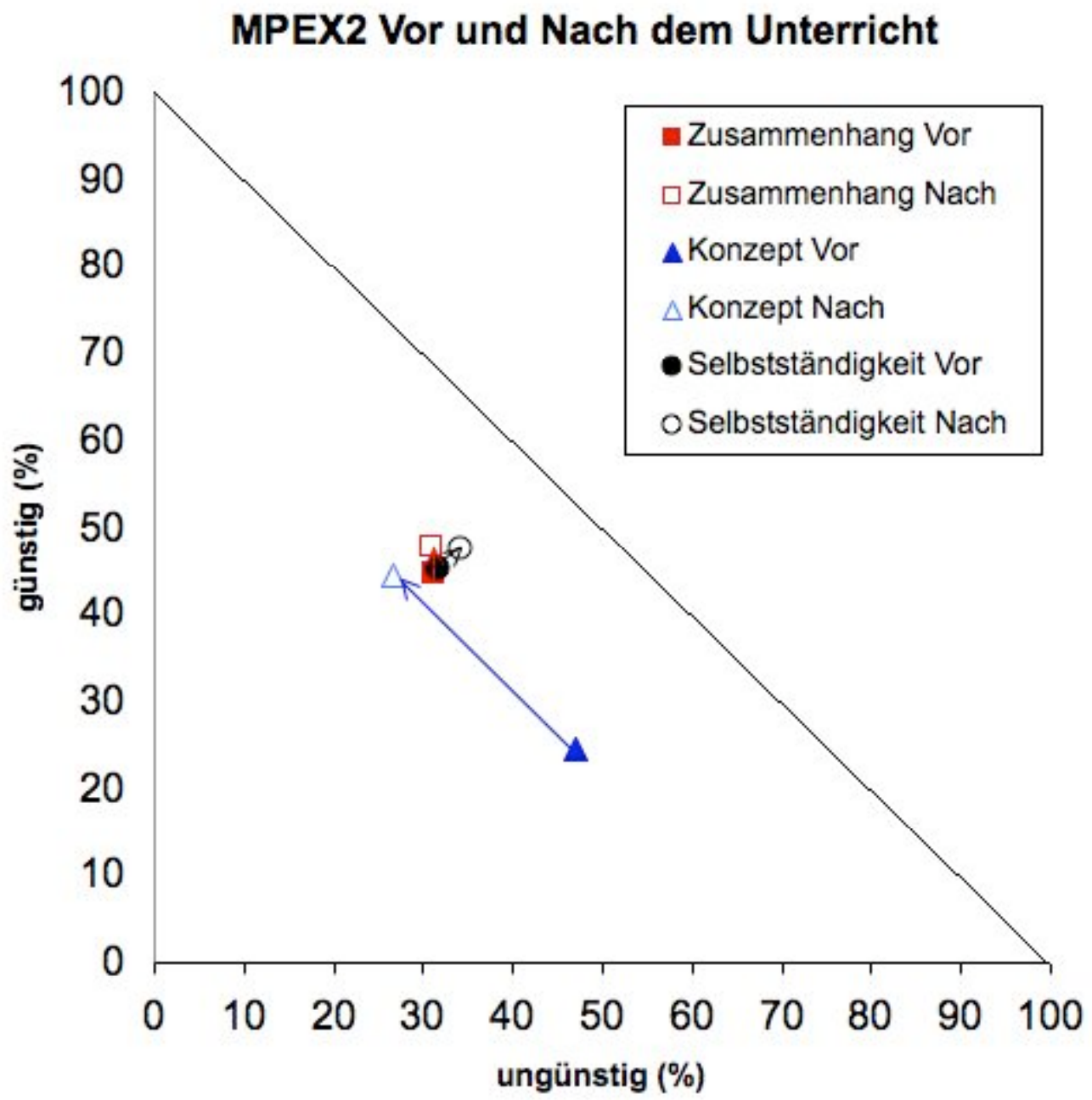

**Abbildung 5:** MPEX2 Resultate aus dem IQP-Kurs. Vektoren zeigen von Vor- zu Nachtests

Ein Strahl quantenmechanischer Teilchen mit bestimmter Energie zielt auf ein Potential, wie rechts zu sehen ist. Die Wellenfunktion für die Teilchen ist unten zu sehen.

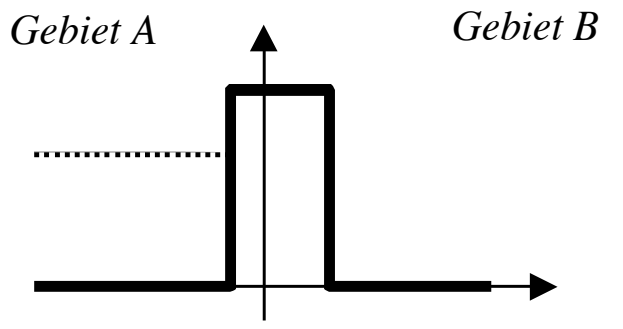

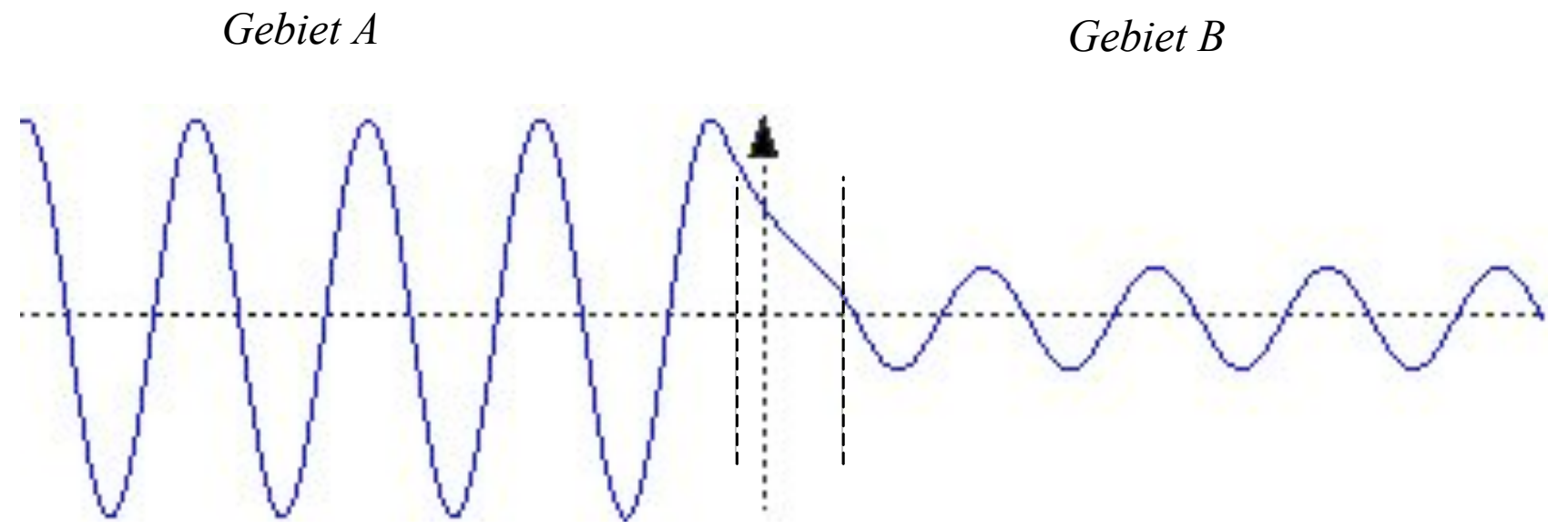

Wie, wenn überhaupt, ändert sich die Wellenfunktion wenn man das Potential

1. breiter macht?
2. höher macht?

Danach fragten wir die StudentInnen, wie sich die Wahrscheinlichkeit und die Energie in Gebiet B ändern im Vergleich zur originalen Situation und im Vergleich zur Energie in Gebiet A.

**Abbildung 6:** Testfrage über das Tunneln von Quantenteilchen

Hauptfach Physik an der Universität Maine geben die richtige Antwort, obwohl sie alle mathematischen Rechnungen durchführen können. Diese Resultate wurden auch an anderen Universitäten gefunden [19]. Wir stellten mehr als 100 StudentInnen die Frage in Abbildung 6. Wir fanden, dass 75% der StudentInnen die Frage über die Energie in Gebiet B richtig beantworten können. Das heißt, die „mathematisch schwachen" StudentInnen des IQP-Kurses antworten besser als StudentInnen mit dem Hauptfach Physik. StudentInnen des IQP-Kurses zeichnen die Skizzen der Wellenfunktion nicht so gut (zwischen 30% und 55% korrekt). Das ist aber überraschend gut für StudentInnen, die am Anfang des Semesters nichts von Wellenfunktionen usw. wussten.

# Diskussion

Wir haben noch andere Belege, dass die StudentInnen im IQP Kurs die Konzepte lernen, aber diese sollen hier nicht weiter beschrieben werden. Der Sinn dieses Aufsatzes ist nicht ein ausführlicher Beweis, dass StudentInnen die schwierigen Ideen der Quantenphysik lernen können. Unser Ziel ist vielmehr, den Kurs detailiert genug zu beschreiben, dass Andere die Ideen in ihren eigenen Unterricht einbauen können.

SchülerInnen der Gymnasien in Deutschland gehen von einer ganz anderen Basis aus als unsere StudentInnen. Das wichtige im IQP-Kurs ist es, dass Lab-Tutorials den StudentInnen die Möglichkeit geben, Konzepte der Quantenphysik selbständig aufzubauen.

## Dank

Die in diesem Beitrag beschrieben Unterrichtsmaterialien wurden mit Hilfe von Katrina Black, David Clark, Roger Feeley, Jeffrey T. Morgan, Eleanor C. Sayre und Richard N. Steinberg entwickelt. Unterstützung kam von National Science Foundation Grants DUE965-2877 und DUE-041-0895, sowie Fund for Improvement of Post-Secondary Education Grant P116B970186.

## Literatur

# Pretest #4

Name:____________________________ Section:______

A. A laser beam is pointed at a photographic plate. The intensity of light from the laser is so low that it only emits one photon at a time.

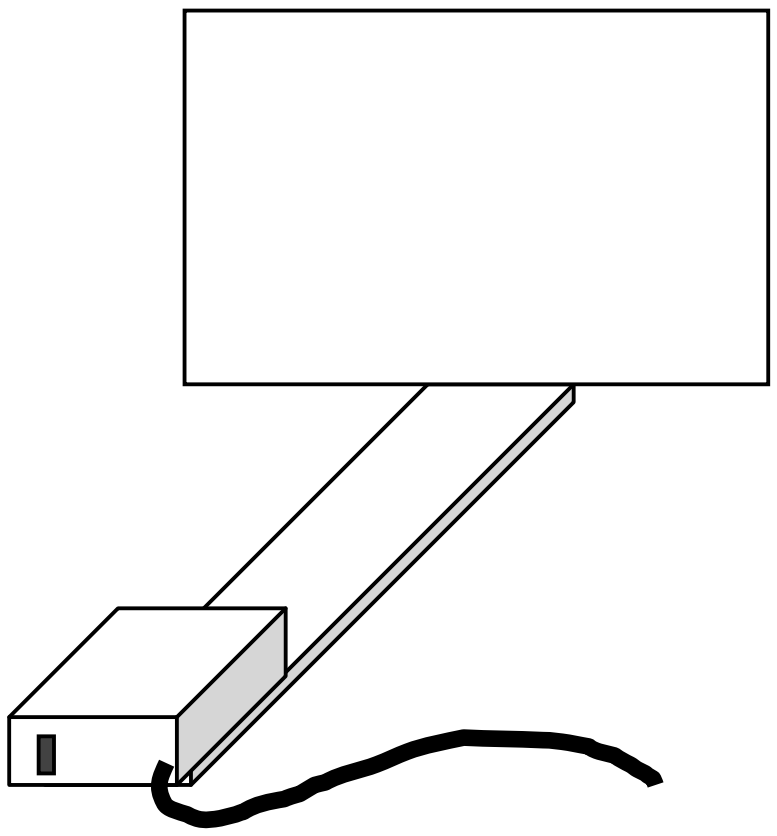

1. Sketch what you would see on the plate after ten photons hit it. Explain how you arrived at your answer.

2. Consider that the laser is allowed to run for a very long time. Sketch what you see on the plate. Explain how you arrived at your answer.

B. Laser light is incident on a slide with two identical narrow slits. A photographic plate is located behind the slide. Again, the intensity of light from the laser is so low that it only emits one photon at a time. What would you see on the photographic plate after ten seconds? Sketch a diagram in the box below. Explain your reasoning.

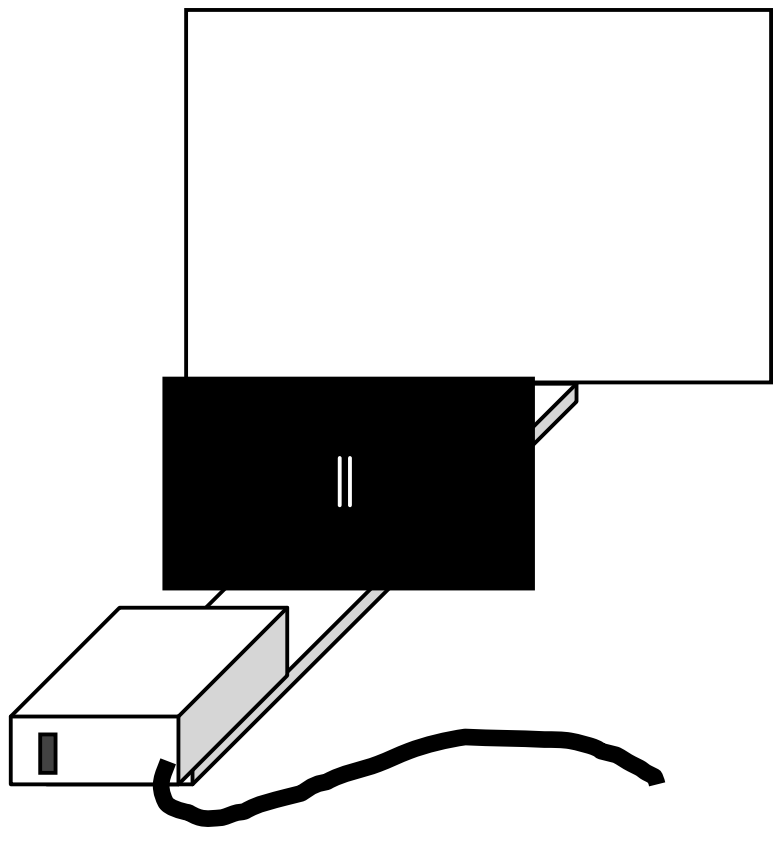

C. A beam of electrons is produced and aimed at a slide with two narrow slits. The intensity of the electron gun is turned down so that it emits only one electron at a time.

1. What would you see on a photographic plate located behind the slide after 10 electrons have hit it? Sketch a diagram in the box below to help account for your answer. Explain your reasoning.

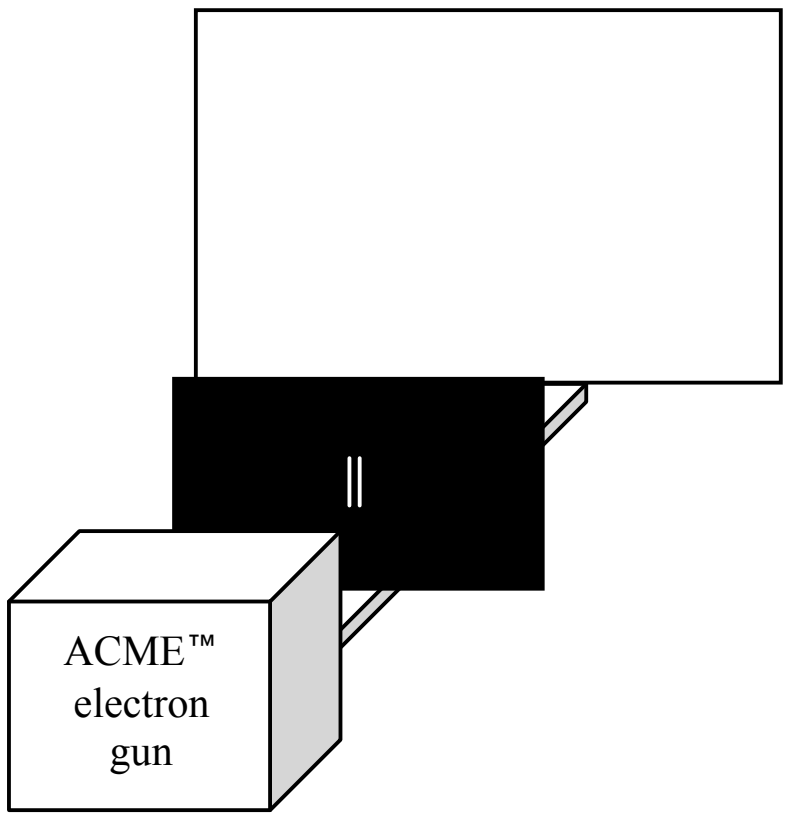

2. The intensity of the electron gun is turned up. What would you see on the photographic plate after a long time? Sketch a diagram, and explain your reasoning.

# *Tutorial #4*: **Doing impossible things**

Name:___________________________

In today's tutorial, we will consider ways in which a single object can do two contradictory things. In order to do this, we first need to be sure that we know what we know about objects. But first, lets kick off tutorial with a board meeting:

## I. Board Meeting #1

Divide your whiteboard in two sections. Title one section "photons" and the other "electrons". Play a word and picture association game – what kinds of things come to mind when you hear these terms? List all the ideas and/or sketches you can think of in 5 minutes on your whiteboard, then gather with the rest of the class to compare ideas.

## II. Thinking about everyday examples of things

A. Imagine shooting a paint ball at a wall (for those who don't know, a paint ball is like a little balloon filled with paint). What would you see?

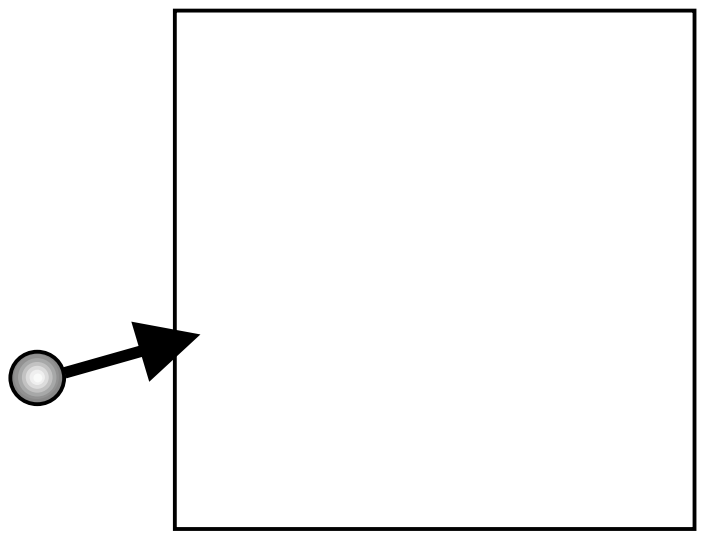

B. Imagine randomly shooting paint balls at a wall with two holes in it. What would happen? What would we see beyond the wall if we had a screen there?

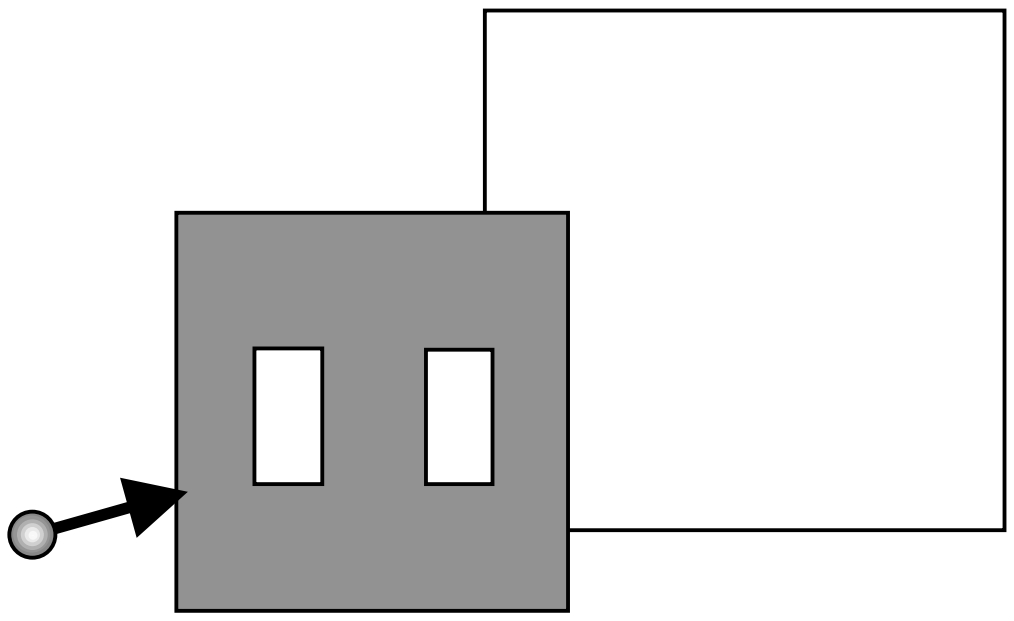

C. Imagine a dyed water wave hitting a gray wall. What do you see on the wall? Is one place different from another? (If you haven't already done so, think about last week's gray barrier.)

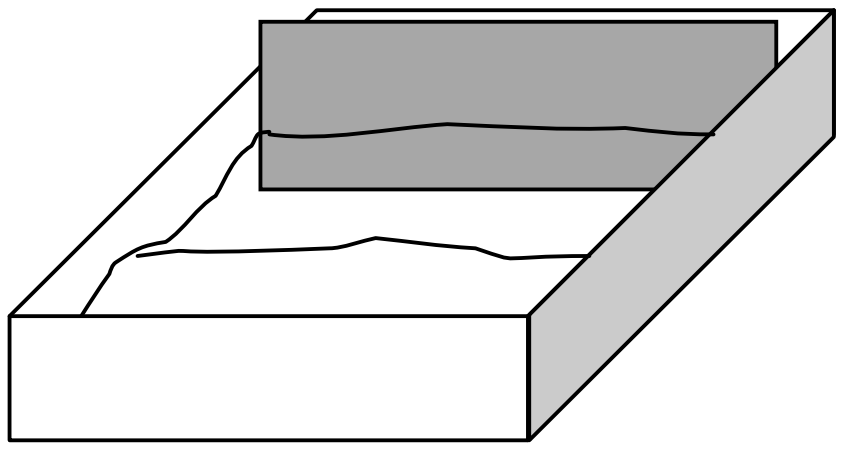

D. Imagine another dyed water wave coming up on a wall with two slits in it. What happens? What would you see on the wall behind? (Yes, we've studied this idea in previous tutorials – be brief in your summary!)

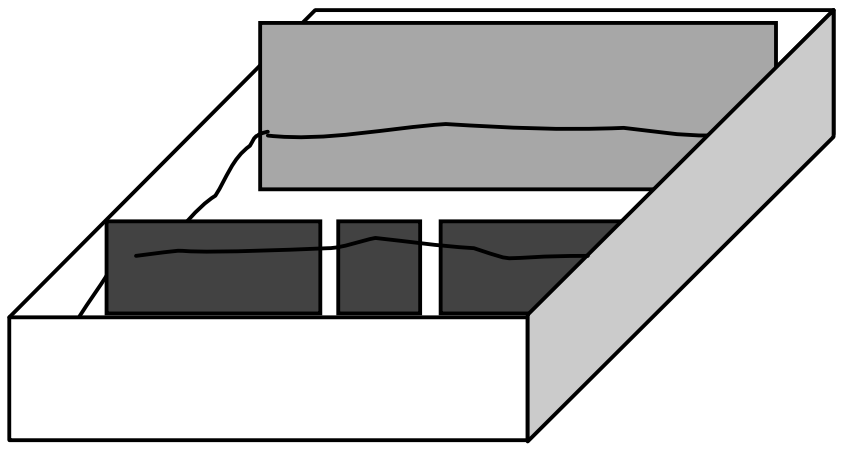

### III. A reminder from last time

It's important for us to remember ideas from last class because we need to keep them in mind while doing new things. We'll just remind you of a few things from before, and then move on.

A. A portion of the pattern produced by a laser beam passing through two very narrow slits has been reproduced below:

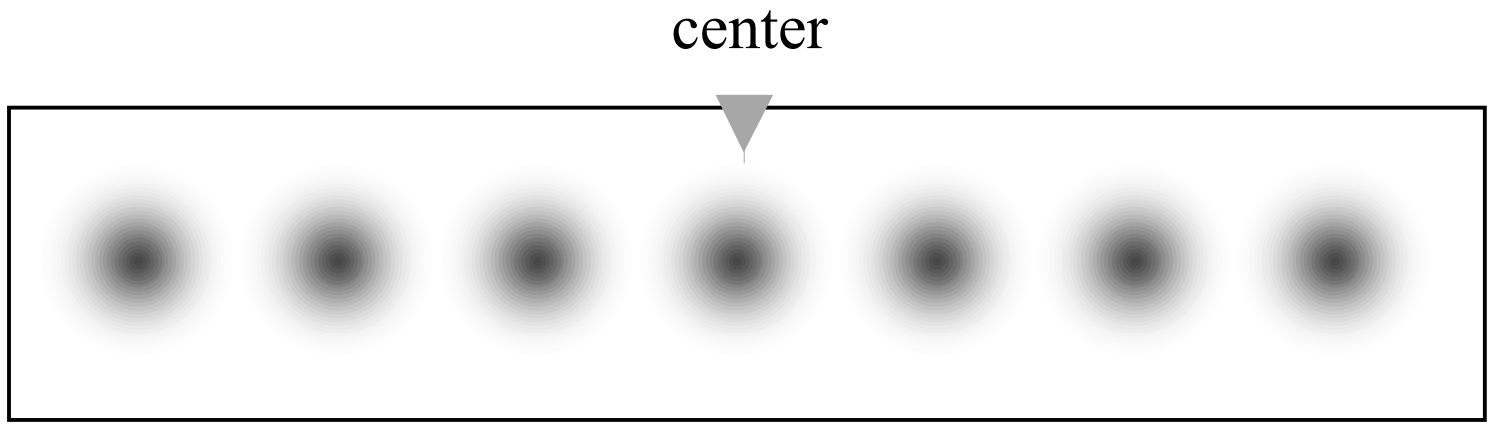

Note the parts of the picture which represent light spots, and recall how we modeled the existence of (1) dark lines and (2) light shining on parts of the screen at which the laser isn't pointed.

B. Suppose the left slit were covered.

1. Sketch what you expect to see on the screen.

2. Your answer should be consistent with the description of water waves passing through a narrow slit from previous tutorials.

   a. Explain what assumptions we make about *light* when we tell you this. This question is designed for you to summarize what you know (a "making my own textbook" kind of question).

   b. Explain what assumptions we make about the *slit* to answer question 1. Again, this is a summary question for your own notes.

C. Compare the pattern seen on the screen when light passes through two narrow slits to the pattern seen in last week's tutorial with the gray barrier when water waves passed through two narrow openings.

1. How, if at all, are the patterns similar?

2. How, if at all, are the patterns different?

### IV. Photons

Another name for a laser is a photon gun. A photon is a small particle of light, and a photon gun like the ones we use in the classroom usually shoots out about, well, a very large number of photons each second (for our lasers in class, it's about 10 million a second). The more photons emitted at a time, the brighter the laser looks. The number of photons that are emitted at a time is related to the *intensity* of the light.

A. Suppose that the intensity of light that the laser puts out is lowered so that it only emits one photon at a time. Suppose the laser shines at a photographic film directly (without a slide in the way). A film turns white when a photon hits it.

1. The big circle represents a normal laser dot. Sketch what you would see on the plate after ten photons hit in the box at the right. How is this like what happens with paint balls?

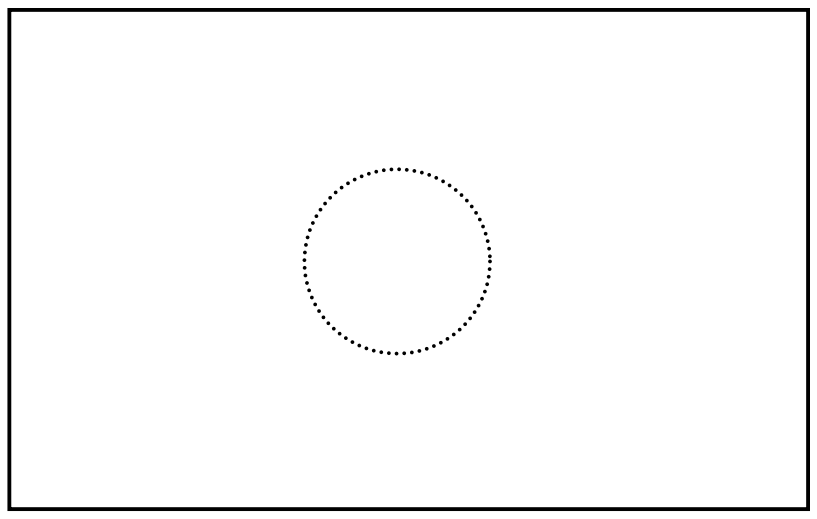

2. On the second box, mark what you would see after a very long time. Explain your reasoning. Is this consistent with the idea of lots of paint balls hitting a wall?

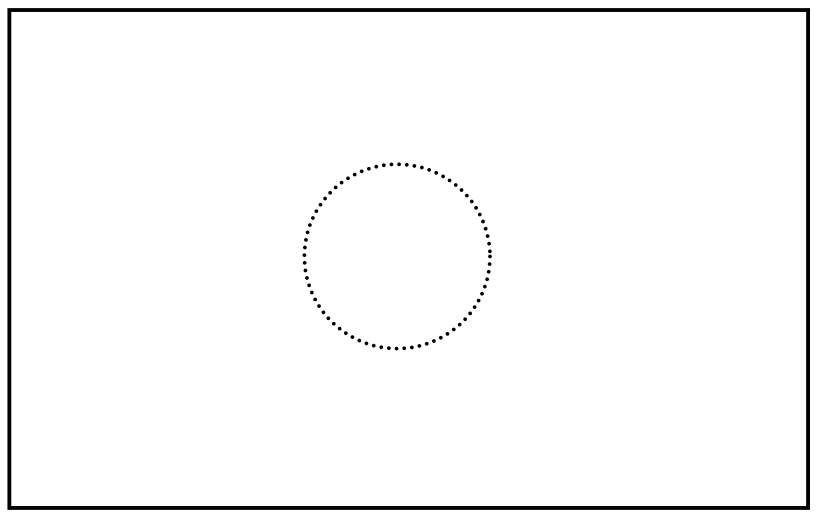

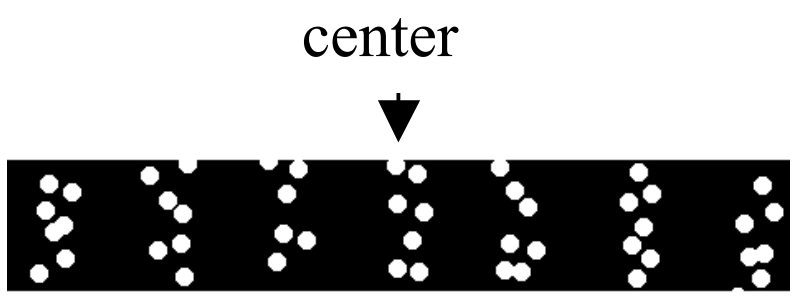

B. The laser light in part A now shines through two slits. The film is located behind the slits. After a short time, a portion of the resulting pattern on the film is shown at right.

1. Is the pattern you observe consistent with the water waves we have studied? Why or why not? Be explicit about the features of the pattern that guide your reasoning.

2. How is the pattern you observe consistent with your description of paint balls on page 1?

C. Suppose one of the slits is covered so that photons can only pass through one of the slits (one at a time). Let's use a fresh piece of film. It might also help to refer to Part III C to make sure you're being consistent with your reasoning.

1. Sketch the appearance of the film after 10 photons have hit. Describe how your sketch differs from your sketch of 10 photons hitting in A1.

2. Describe how, if at all, the appearance of the film would change after many photons hit.

3. Would you see the dark bands? Keep in mind that many photons hitting is the same as hitting a laser turned up at regular intensity.

## V. Electrons

A. An electron is like a very, very small paint ball. Without the paint. Like paint balls from paint ball guns, we can shoot electrons from electron guns. Your television, if it's "old school" and not one of those new-fangled flat screens, has an electron gun in the back. It shoots electrons at the screen which glows when an electron hits it. Suppose that an electron gun shoots out one electron at a time, and those electrons hit a photographic plate. A photographic plate is like film, and it lets us see where electrons hit.

1. The rectangle on the right represents the plate. Sketch what you would see on the plate after ten photons hit it. How is this like what happens with paint balls?

2. On the second box, mark what you would see after a very long time. Explain your reasoning. Is this consistent with the idea of lots of paint balls hitting a wall?

B. The beam of electrons is directed at two slits. A photographic plate is located behind the slits. After a short while, the electrons make a picture like the one on the right. (Note: This picture and ones in a subsequent section come from an actual experiment, not a computer simulation.)

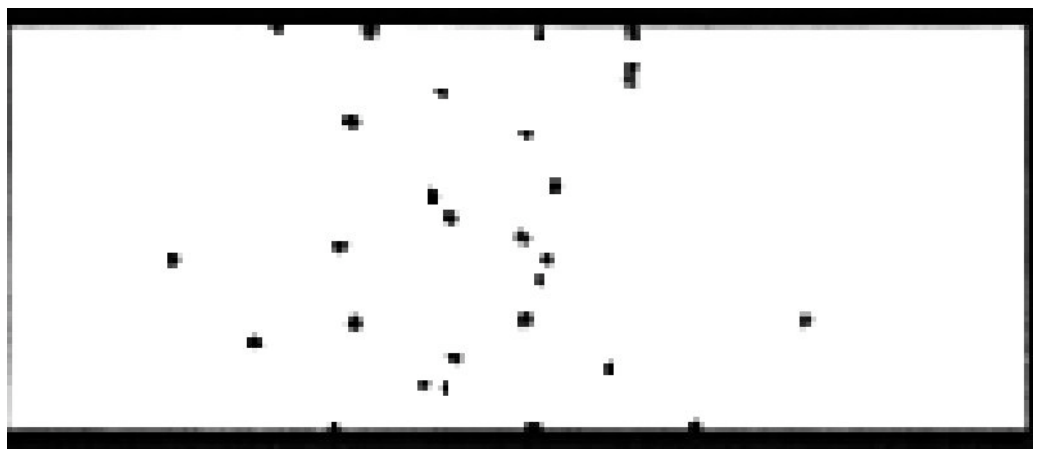

1. Do you see a pattern in the dots created by the electrons?

2. We've said that electrons are like paint balls. In this scenario, we're shooting electrons at a 2 slit mask, just like we did with paint balls in II.B. Does this picture look similar to what you predicted in I.B.?

## VI. Board Meeting #2

So far today you've talked a lot about particles – paint balls, photons, and electrons – and a little about waves. Your instructor will assign you one of the questions below to address as a group.

1. How, if at all, are your answers about the beam of electrons directed at two slits (Part V) similar to your answers about a small number of photons going through two slits in Part IV?

2. Would paint balls do what electrons do?

3. Consider that the low intensity laser only emits one photon at a time. Why don't those photons ever land in the dark stripes?

Discuss the questions in class. Arrive at a consensus on each, and then move on to the next part of the tutorial.

### VII. Reinterpreting Electrons OR Electrons are not really paint balls.

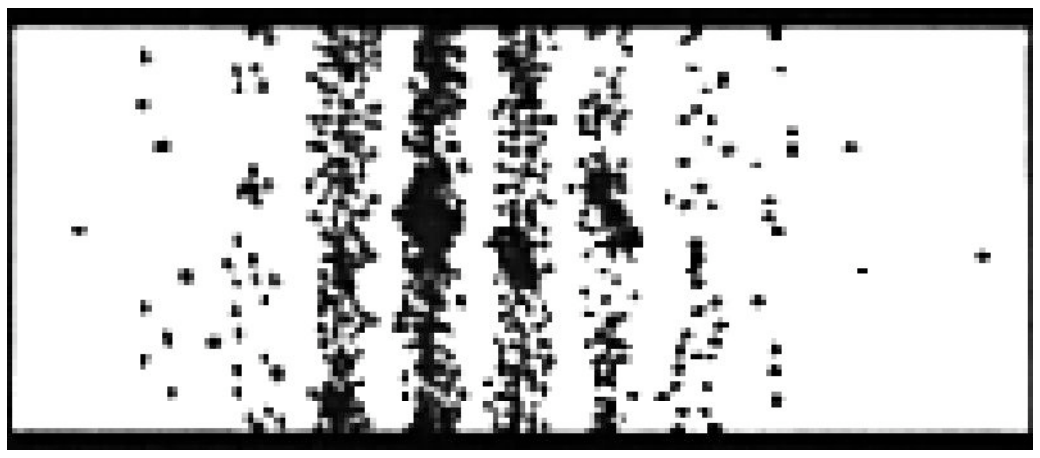

While you were at the board meeting, the experiment of electrons being shot at the two-slit mask continued. (Just play along with us on this, thanks…)

A. Upon returning to your tables, a second photograph of the screen is taken.

1. How is this picture similar to the sketch you made in V.A?

2. How is it dissimilar?

3. Discuss what aspects of this experiment are consistent with electrons acting like a particle.

B. Imagine that we created an experiment like in part A and divided the screen into bins about a quarter of an inch wide and as tall as the screen is. A graph of how many electrons hit in each bin is called a histogram.

1. Please make a histogram of electron hits below.

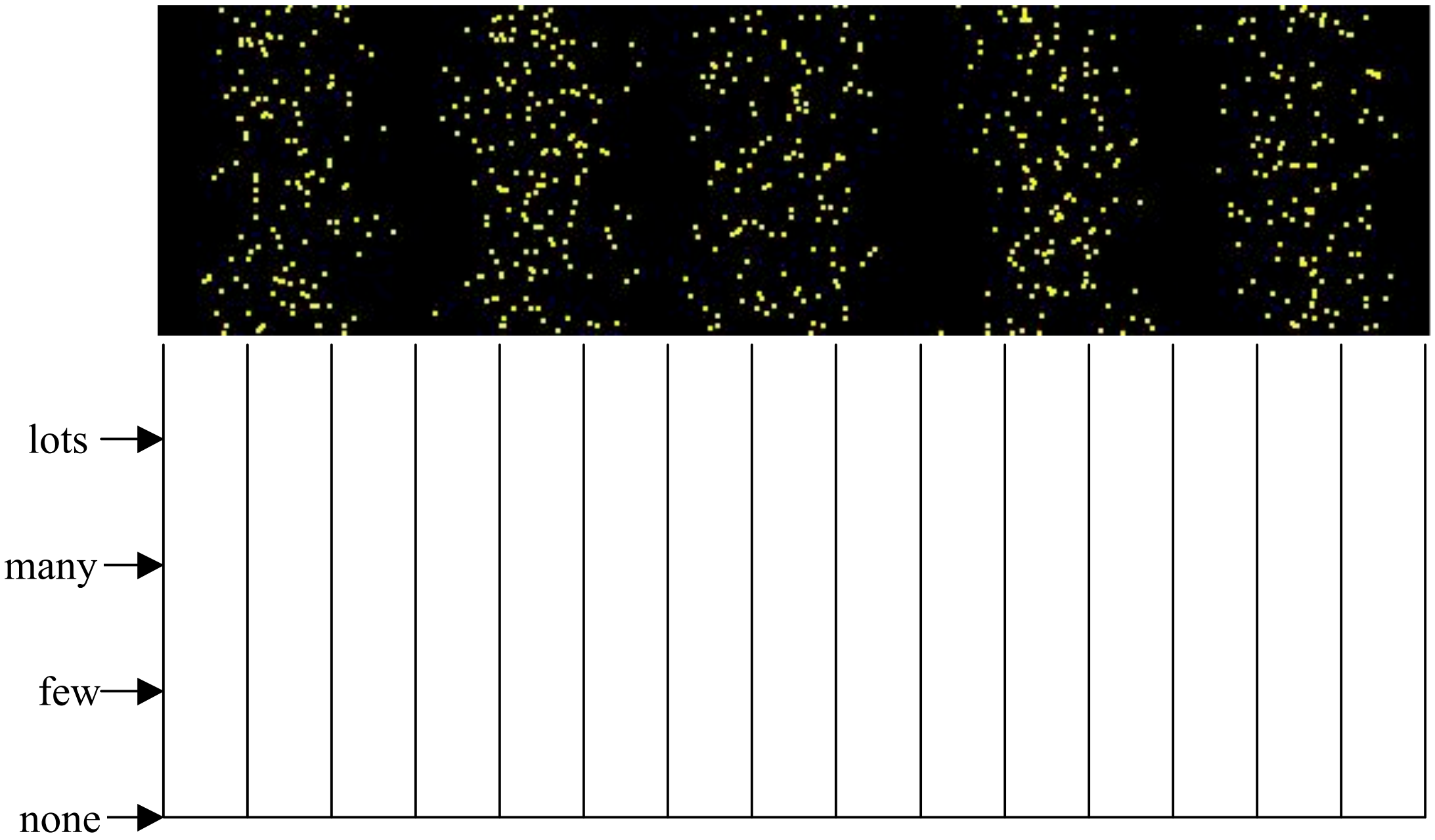

2. As time goes by, more and more electrons hit the screen. Predict how the histogram will look a long time later, and explain your reasoning.

## VIII. Using a computer to help take data

It's not only tedious for you to guess-timate the values of your histogram, and it's not terribly accurate, either. Computers are good for tedious but accurate work. For instance, a program could make the bins very narrow, and is most likely very good at counting quickly.

Open the file titled "screen.html" on your desktop and wait for the applet to load. It might take a bit. The URL is: http://perlnet.umaine.edu/QuantumMechanicS/en/version1/p0000a.htm

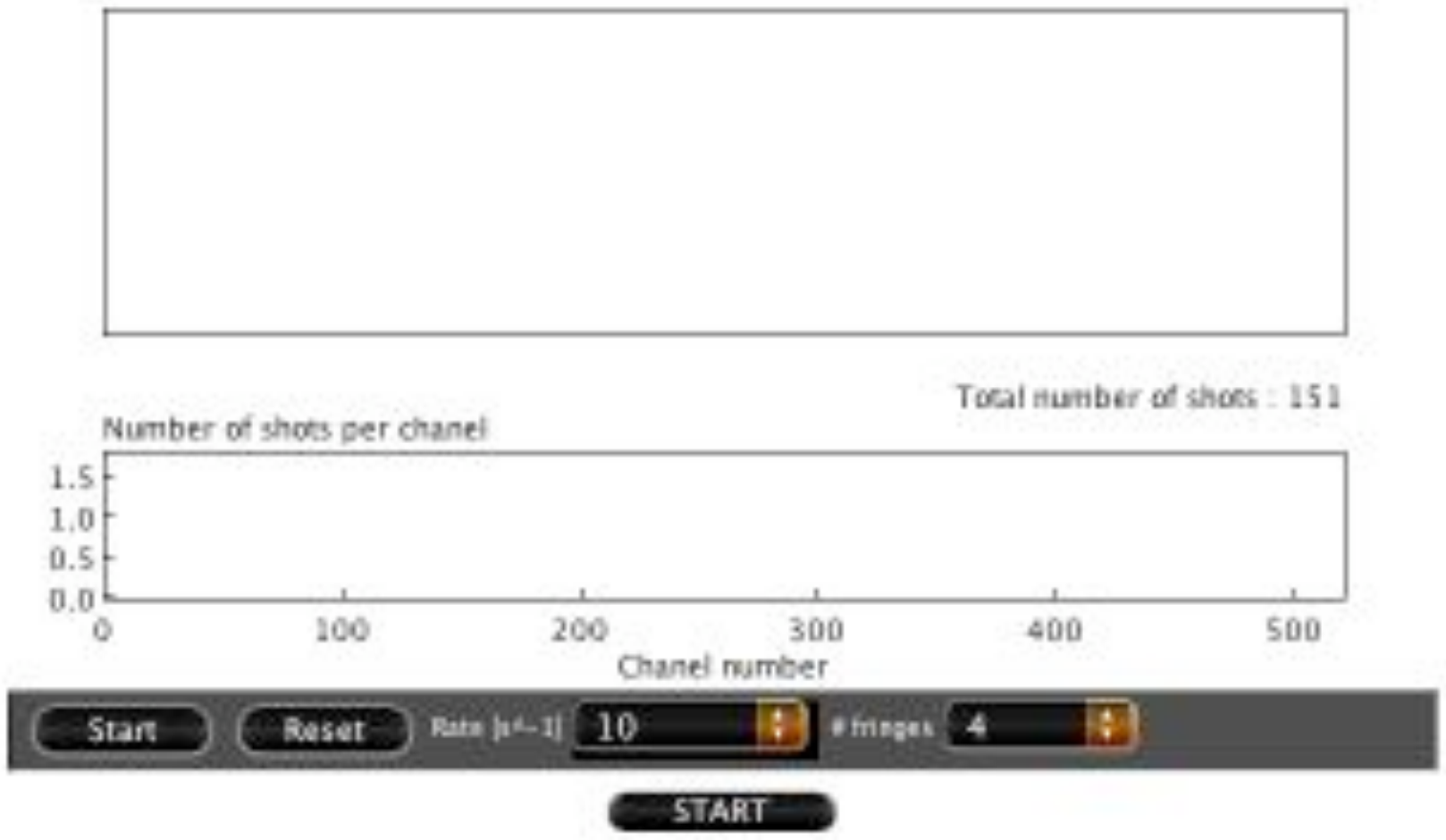

**Upper part**: a screen. Electrons that hit the screen make yellow dots; those dots stay around until you clear the screen by hitting "reset" in the controls.

**Lower part**: Histogram, which counts how many particles have hit a given location.

**Controls**: The "Rate [s^-1]" menu controls how many dots per second appear (set at 10 in this screen capture). The "# fringes" menu controls how many light bands will show up on the screen. You can also pause and reset the applet. If these controls are greyed-out, click the stop button in the middle near the bottom. Notice that there are *two* Start buttons. The center-bottom one starts the applet itself, the left control one determines whether or not electrons are being shot at the screen. (Sorry it's confusing, we didn't write it.)

Set the applet to make 5 fringes, and set the rate to 100 dots/second. Click reset to clear the screen. Observe the histogram as time passes.

A. How, if at all, is it like the work you did in part B on the previous page?

B. The shape of the histogram might look familiar, in that it's a lot like what you said the gray barrier would look like. At the same time, there are lots of differences between the two.

   1. What are the differences?

   2. What is the most important feature that both the gray barrier and the histogram show?

C. Let the simulation go for a longer time. The screen could look like the picture below.

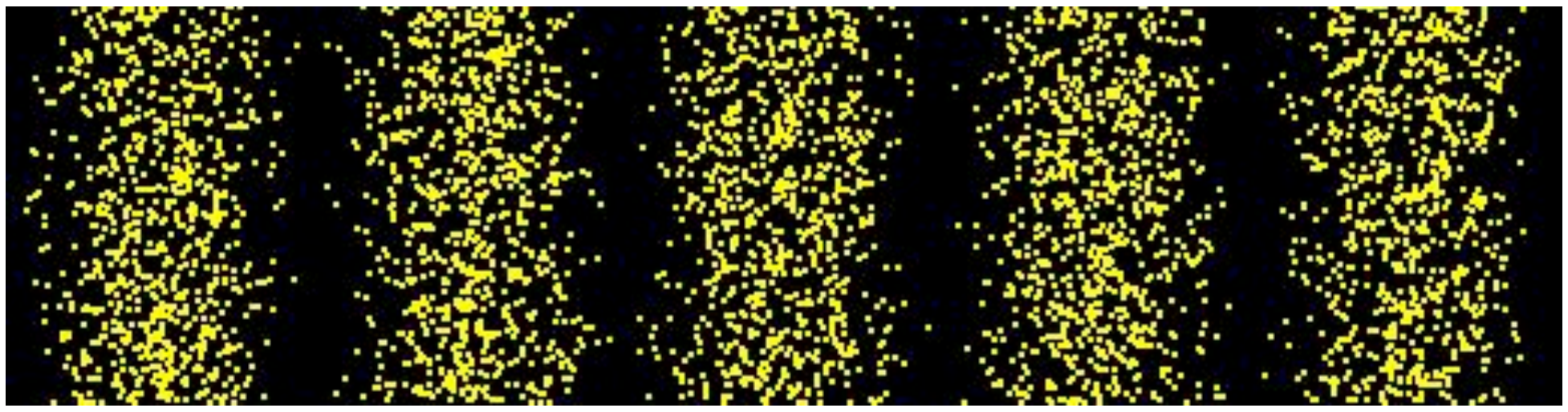

1. Is this what you would have predicted? Resolve any discrepancies.

2. In the picture below, you can see the results of letting the two-slit electron experiment from sections V and VII of this tutorial continue. Compare the three images from the experiment and describe what features become more noticeable as time passes.

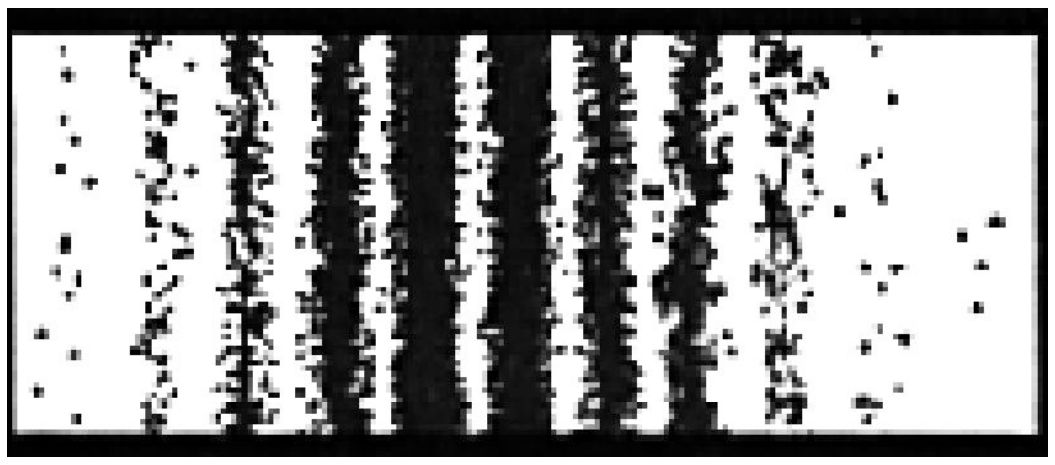

To test our prediction, we can let the simulation run until the end of class. Do that now…

3. Compare what we see on the screen and what we see on the histogram. Why does the histogram show nice curvy lines, while the screen shows washed out yellow? What does this say about the "photographic plate" that is on our screen?

D. We can use our observations to figure out certain properties of electrons.

1. How, if at all, is it appropriate to say that electrons behave like waves? What must we observe to conclude that a thing is like a wave?

2. How, if at all, is it appropriate to say that electrons behave like particles?

E. Suppose the experiment in part A is repeated except that one of the slits is covered so that electrons can only pass through one of the slits.

1. Sketch how, if at all, the appearance of the screen would change. Explain your reasoning.

2. Is your answer consistent with your description of light going through one slit? If not, resolve the discrepancy.

## IX. Finding consistencies

In this lab, we've studied paint balls, electrons, and photons, and we've reminded you of water waves. The questions in this section are a place for you to summarize the ideas developed in this lab. You may need more paper than we provide! This should form a core part of your class notes.

A. In what ways do electrons behave like paint balls? Support your answer by description and with reference to specific experiments.

B. In what ways do electrons behave like water? Support your answer by description and with reference to specific experiments.

C. In what ways are electrons like photons? We should tell you that our experiments haven't given us enough information to say how the two are *different*. Brainstorm with your group: what experiments might you carry out to distinguish between the two?

## X. Board Meeting #3

Electrons and photons seem to do impossible things. They act like waves *and* they act like particles. More to the point, sometimes they *don't* act like particles, and sometimes they *don't* act like waves. And when we're not paying attention, we don't know which they are acting like.

We're going to explore this idea more in this board meeting, and try to pull together all the analogies we've made (to paint balls, water waves, and more).

1. In this class, we have used rays to describe light and water. When we use rays, do we mean waves, particles, or both? [It may help to draw some pictures. What is moving in a light ray? What, if anything, is moving in a water ray?]

2. It bothers some people that electrons can act both like waves and like particles. Does it bother you? Why might it bother people?

# *Tutorial Homework #4*: **Doing impossible things**

PHY 105 – *Descriptive Physics* – Fall 2005 Name:___________________________ Section:______

A. Suppose a beam of electrons is aimed at two slits in a slide placed in front of a screen. After a short time the screen looks like the one at the right.

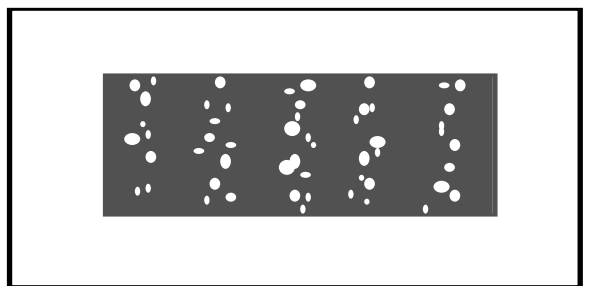

1. How, if at all, would the picture change if you decreased the spacing between the slits? Explain which analogy you are using in answering your question.

2. How, if at all, would the picture change if you moved the screen closer to the slide containing the slits? Again, explain which analogy you are using in answering your question.

B. Compare your answer to A to your descriptions of water and light waves. For each, say how, if at all, they are similar *and* different:

1. water waves:

2. light waves:.

C. What evidence does the picture in part A give that electrons act like particles? Like waves?

D. Below and to the right is a top view of an electron beam source, a slide with two narrow slits, and a screen as described in problem A. The pattern on the screen (not shown) is like that in problem in problem A in that there is an interference pattern.

**Diagram not drawn to scale.**

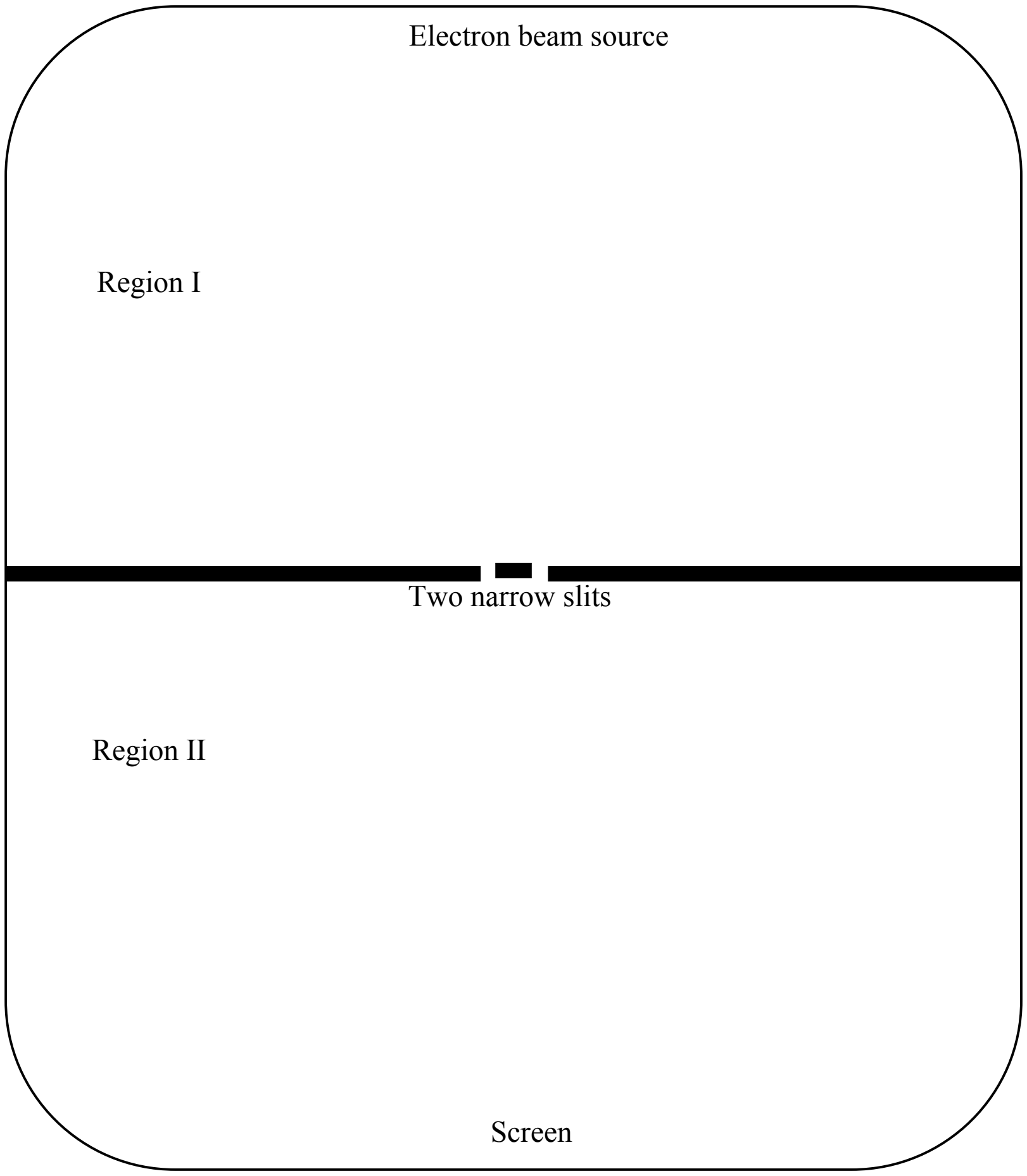

1. Draw the *wave pattern* for the electron beam in regions I and II. Explain your reasoning for each region.

2. Use a different color ink to draw the *ray diagram* for the electron beam in regions I and II. Explain your reasoning for each region.

3. Are the wavelengths of the electron beam in regions I and II different or the same? What analogy are you making when answering this question?